# The Recall Trap: A Recall-Maximizing Retriever Configuration Reduces Issue Resolution in Fixed-Budget Code Context

Alexander Adkins, Teimuraz Trapaidze

*Make Games Initiative*

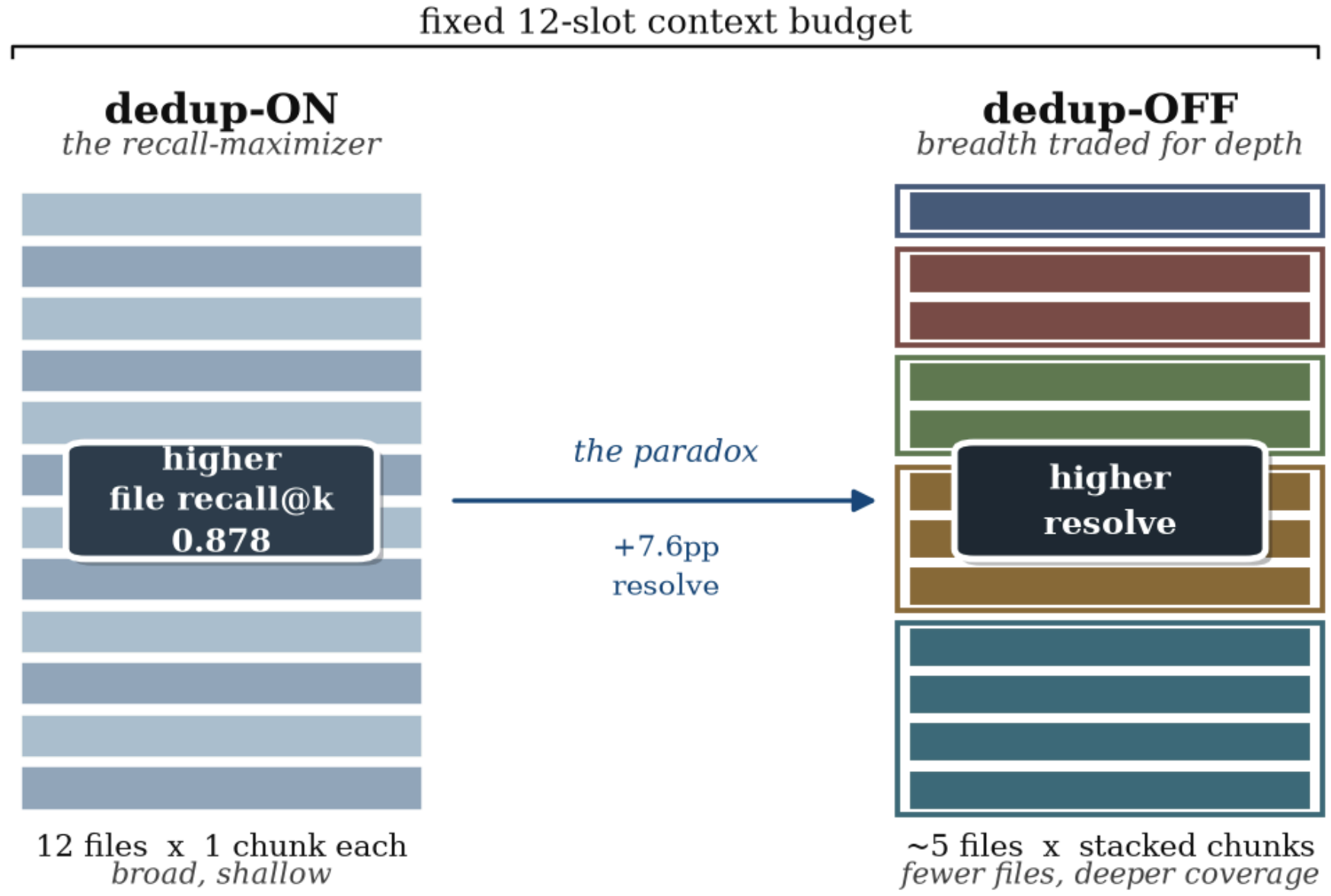

**Figure 1.** The recall trap at a fixed 12-slot budget. **Dedup-ON** (left), the recall-maximizer, spends the budget on ~12 distinct files with one shallow chunk each, maximizing the chance the gold file appears somewhere in the pack (gold file present in 0.878 of packs). **Dedup-OFF** (right) spends the same 12 slots on ~5 files with multiple stacked chunks each, trading file breadth for within-file depth. The paradox: dedup-ON wins file recall@k yet dedup-OFF resolves more (gpt-5.6-sol 46.8% vs 39.2% single-shot, +7.6pp, p=0.0003).

## Abstract

Retrieval components for code assistants are tuned against retrieval metrics: a configuration that raises recall@k is adopted, and downstream task success is assumed to follow. We report a controlled case study in code repair, not a new phenomenon but a deployed-flag, execution-graded instance of the known relevance-diversity and objective-mismatch tradeoff (Levy et al., 2025). On SWE-bench Verified we inject a retriever's hits as a fixed 12-slot context pack with no search tools and toggle one flag (one-chunk-per-file deduplication) on an otherwise identical stack. The flag is the higher-recall configuration (gold file present in 0.878 of served packs against 0.806 disabled), yet disabling it, trading file breadth for within-file depth, raises the single-shot resolve rate: gpt-5.6-sol +7.6pp (39.2% to 46.8%, n=500, McNemar exact p=0.0003), and a pre-registered open-weights replication any reviewer can re-run (Qwen3.6-27B, +3.6pp, n=499, p=0.0133); both survive repository-clustered inference. The gain tracks within-file anchor dose, and a random-chunk control refutes an argmax-selection artifact. We map where it holds: it reverses on a lexical BM25 retriever (−3.2pp, significant cross-paradigm interaction), is not detected under unrestricted-Read agents (a powered null), and across four languages (SWE-PolyBench, N=617) is positive but not significant (+2.6pp, p=0.056), a mapped boundary rather than a confirmed extension. Operationally, at a tight fixed budget: do not hard-deduplicate by file, and A/B packing policies against the task, not the metric the flag was tuned to.

## 1. Introduction

Modern code-assistance systems increasingly deliver repository context to a language model as a pack: a fixed number of retrieved snippets injected into the prompt, with no further search. This regime, fixed-budget and pre-injected, is deployable and appears in pack-style RAG serving, in seed contexts for constrained agents, in cost-capped batch pipelines, and in local models that cannot run agentic loops. Latency, cost, and context-window pressure cap how many slots the retrieval layer may fill, and every slot given to one file is a slot taken from another. In this regime, retrieval is zero-sum. We study exactly this sole-context single-shot regime; we do not claim it describes interactive assistants that add context in a loop (Section 5.5 shows the effect does not survive there).

Retrieval components for such systems are tuned against retrieval metrics. In the system we study, a one-chunk-per-file deduplication flag was adopted because it raised file-level recall@k in the retrieval index eval from 0.666 to 0.817. Deduplication of this kind is standard, and its logic is sound for the metric: if the budget is k slots, spending two of them inside the same file cannot improve the chance that the gold file appears somewhere in the pack, so diversifying across files improves file-level recall. Diversifying a ranked list to raise coverage is the classic maximal-marginal-relevance heuristic (Carbonell and Goldstein, 1998), and diversity-aware IR objectives such as α-nDCG (Clarke et al., 2008) and subtopic diversification (Santos et al., 2010) formalize the same one-per-cluster preference. File-dedup is a hard, file-keyed instance of that family. We do not claim file recall@k is the only retrieval metric in use; it is one recall-justified production default, and rank-aware or diversity-aware metrics would score the two arms differently.

This paper documents what that one tuning decision does to the downstream task. We hold the retriever, ranking, embedder, 12-slot budget, and grading harness fixed and toggle only the dedup flag. The primary evidence is single-shot: one API call per instance, no tools, official SWE-bench Docker grading, empty patches counted as failures.

- gpt-5.6-sol (n=500, single-shot): dedup-ON 39.2% (196/500) against dedup-OFF 46.8% (234/500), Δ +7.6pp, discordant 73 against 35, McNemar exact p=0.0003.
- DeepSeek-v4-pro (n=499, single-shot): Δ +6.2pp (raw), discordant 53 against 22, McNemar exact p=0.0004; supportive evidence carried by a producibility channel (Sections 4.5, 5.2).
- Qwen3.6-27B (n=499, single-shot, open weights, pre-registered replication): dedup-ON 9.2% against dedup-OFF 12.8%, Δ +3.6pp, discordant 33 against 15, McNemar exact p=0.0133, conditional p=0.035 (the open-weights confirmatory arm; the weak-open Qwen3-Coder-30B-A3B anchor gives +4.2pp at K=12 and a monotone budget×dedup trend, Section 5.9).
- DENSE-1, gpt-5.6-sol on dense-only packs (n=494, single-shot): Δ +5.7pp, discordant 69 against 41, raw p=0.0097, conditional p=0.0225 (the retriever-generality control).

The configuration that maximizes file recall@k reduces the resolve rate at this fixed budget. We call this the recall trap: at a zero-sum context budget, the diversity-maximizing packing policy prefers shallow breadth (about 12 files, one chunk each) over deep coverage of fewer files (about 5 files, multiple chunks each), and the solver pays for it. We position this not as the discovery of a new phenomenon but as a controlled, causal, code-repair case study of a known objective-mismatch and relevance-diversity tradeoff (Section 3), together with a mechanism-linked map of where recall-maximization is safe versus harmful.

The closest prior mechanism is Levy et al. (2025), who hold total context length and gold position fixed in multi-hop QA and vary only the number of documents, a controlled manipulation that isolates a degradation of up to about 20% from breadth alone, with a Qwen exception that pre-figures the model-dependence we observe. Their treatment isolation is in fact cleaner than ours, which is a compound flag (Section 2); we do not claim to be the first causal demonstration of the phenomenon. We differ on three axes a reviewer will ask about. Our manipulation is a standard, recall@k-justified retrieval configuration flag rather than an experimenter's document-count dial, so the harm arises from a packing policy a practitioner would plausibly adopt for its recall gains; our task is execution-graded repository repair rather than QA answer accuracy; and we show the effect does not extend to a lexical BM25 retriever, where it reverses (Section 5.3), so it is not a universal document-count law. We contribute this domain, this instrument, and the boundary map, not the qualitative claim that recall can mislead, which this prior art and the diversity literature already establish (Section 3.1). We say the result is more than a mere code instance of Levy et al. in this specific sense, and nowhere claim a new phenomenon.

### Contributions

1. A deployed-flag, code-repair instance of the fixed-budget breadth-versus-depth tradeoff. We show that a standard, recall@k-justified retrieval configuration flag (one-chunk-per-file file deduplication) reduces execution-graded issue resolution at a tight fixed budget, measured paired at n=500 and surviving repository-clustered inference, on two confirmatory model arms (gpt-5.6-sol +7.6pp; and a pre-registered replication on open weights, Qwen3.6-27B +3.6pp, p=0.0133, independently re-runnable; both survive a conditional estimand) with a third supportive arm carried by patch producibility (DeepSeek-v4 +6.2pp, conditional null). The effect holds across a roughly 5x resolve-capability span (weak-open Qwen3-Coder-30B-A3B +4.2pp, strong-open Qwen3.6-27B +3.6pp, frontier-closed gpt-5.6-sol +7.6pp, all at K=12). The novelty over the nearest prior (Levy et al., 2025, which is itself a controlled manipulation, and DeepDiscovery, which is paired and execution-graded on the same benchmark) is not that the inversion is causal or paired but that it arises from a real recall-tuned packing policy, in the tight zero-sum budget regime, on repository repair. We do not claim it differs in kind from the known diversity, granularity, and objective-mismatch tradeoff, and we hedge it to this single-flag compound instrument and the tested regime.
2. A mechanism-linked boundary map: where the effect does and does not appear. The effect holds for two embedding retrievers (fusion, and the dense DENSE-1 replication whose clustered interval includes zero), does not extend to a lexical BM25 retriever, where it reverses (dedup-OFF minus dedup-ON −3.2pp) with a significant cross-paradigm interaction, tracks within-file anchor dose (a supported but partial mediator; BM25's depth arm raises it only +0.029 against fusion's +0.089), concentrates on single-file fixes, and does not appear under unrestricted-Read agents (a powered null). A pre-registered multilingual test (SWE-PolyBench, four languages, two models, N=617) is directionally positive but not significant (+2.6pp, p=0.056), a mapped boundary whose cross-language heterogeneity tracks the instance pool rather than the surface language, not a confirmed extension. Generality beyond the two embedding retrievers tested is not established; the BM25 reversal shows the effect is not a universal document-count law, and its co-variation with anchor dose across retrievers is the most transferable output, but we do not claim a clean retrieval-paradigm boundary from one BM25 configuration.
3. A controlled refutation of the argmax chunk-selection alternative. A random-chunk-per-file control (decision rule fixed in a dated design note before any generation; the note carries no pre-data commit, so we claim no pre-registration for it; see Appendix B) refutes the alternative that the flag keeps the wrong within-file chunk: random reselection of a non-argmax chunk from the same files is worse than the argmax the flag keeps (3.0% against 6.6%, against depth 12.4%). This rules out an argmax-selection artifact ("the flag keeps the signature chunk, not the body"). It does not by itself separate within-file depth from the accompanying change in file set or from same-file contiguity; a file-set-fixed depth control and a contiguity control are the remaining identifiers and are future work.

### What we do not claim

We do not claim a numeric optimal file count; a discovery of a new phenomenon rather than a controlled case of a known one; separation of the compound treatment (files against depth against distractor removal); an inference-time surrogate for the oracle-defined anchor dose; retriever-agnosticism, nor a clean retrieval-paradigm boundary (the effect holds for the two embedding retrievers tested and reverses for the one lexical BM25 configuration tested, which is not the same as establishing generality across paradigms); token-level budget matching (depth packs are 5 to 6% larger in tokens, addressed as a covariate in Section 6.2); applicability to tool-using search agents (the effect is not detected there), or to larger budgets, other embedders, or other chunkers not tested; confirmed multilingual generality (the pre-registered pooled multilingual test was directionally positive but did not reach significance, Section 5.10); or generality beyond SWE-bench-Verified-like popular public Python repositories.

---

## 2. Background and Problem Setting

The fixed-budget pack regime. A retrieval layer answers an issue description with a ranked list of code chunks, and the top-12 chunks are rendered into the solver's prompt. In the primary single-shot regimen the solver must emit its full patch in one completion from the pack alone. This isolates retrieval configuration as the only manipulated variable.

The dedup flag and what it confounds. With file-dedup ON, the ranking is post-filtered to at most one chunk per file before the top-12 cut, so the pack spans about 12 distinct files with one chunk each. With file-dedup OFF, the raw ranking fills the slots: typically about 5 files, with multiple chunks per file wherever the retriever ranks several chunks of the same file highly. The candidate pool and scoring function are identical across arms; the flag reallocates slots. This reallocation is a compound treatment, not a single clean variable: because dedup-ON post-filters before the top-12 cut, it substitutes deeper-ranked cross-file chunks for shallower same-file ones, so the arms differ in served rank and score distribution, in the number of files, in chunks per file, and in which within-file chunk is kept, not only in slot allocation. We name this compound explicitly and, in Section 5.4, use a dedicated control to rule out the within-file-selection component (which chunk is kept) as the cause; we do not claim to isolate within-file depth from the accompanying change in file count, which would require a file-set-fixed control we did not run. We do not claim the arms share an identical ranking.

Why dedup is the recall-maximizing configuration. File-level recall@k asks whether the gold file appears anywhere in the k slots. Spending a second slot inside a file already in the pack cannot help that metric, so dedup-ON is near-optimal for it. This holds both in the retrieval index eval (0.666 to 0.817) and, re-derived on the packs the models actually saw, on the served packs (gold file present in 0.878 of dedup-ON packs against 0.806 of dedup-OFF packs, n=500). The recall-maximizing premise is therefore not an artifact of the index eval; it reproduces on the served packs.

---

## 3. Related Work

We position this work as a controlled causal case study, in code repair, of a tradeoff established across several literatures: relevance-diversity objectives in IR, objective-mismatch between retrieval metrics and downstream generation, fault-localization granularity in program repair, and retrieval-granularity choices in RAG. We engage each below and mark, in each case, what is prior and what our experiment adds.

### 3.1 Diversity objectives and the QA/IR prior art we build on

The preference for one item per cluster, which file-dedup enforces at file granularity, is the core of diversity-aware retrieval: maximal marginal relevance (Carbonell and Goldstein, 1998), α-nDCG and novelty/diversity evaluation (Clarke et al., 2008), and query-aspect diversification such as xQuAD (Santos et al., 2010). These objectives trade per-item relevance for coverage of distinct aspects, exactly the breadth-over-depth trade our flag makes at file granularity. That optimizing a coverage/diversity objective can reduce task utility when the consumer needs depth is therefore not a new phenomenon; it is the objective-mismatch that this family already anticipates.

The same direction is documented for readers of packed context. Cuconasu et al. (Power of Noise, SIGIR'24) showed that retrieved distractors damage QA readers; their secondary "random noise helps" direction has since been attributed to an experimental artifact (Mazuryk et al., 2026). Liu et al. (Lost in the Middle, TACL'24) showed positional degradation as contexts grow. Most directly, Levy et al. (2025) hold total context length and gold position fixed in multi-hop QA and vary only the number of documents, isolating a degradation of up to about 20% from breadth alone, with a Qwen exception that pre-figures our model-dependence; this is the depth-over-breadth mechanism cleanly isolated, and we treat it as the nearest prior mechanism (Section 1). Bala (2026) argues that recall is the wrong retrieval metric when the reader consumes a packed context, in multi-hop QA. Salemi and Zamani (eRAG) showed that standard retrieval metrics only weakly predict downstream generation quality. Our result is a code-repair instance of this objective-mismatch, contributed as a controlled causal toggle of a deployed default, execution-graded, with an embedding-retrieval boundary that a universal document-count reading would not predict. We do not claim it differs in kind from this prior art.

### 3.2 Fault-localization and program-repair granularity

The failure mode our mechanism turns on, gold file present but gold lines absent from any chunk (Section 5.7), is the central concern of LLM fault localization and automated program repair, where hierarchical file/method/line localization and the quality of the localized region drive repair success. Agentless (Xia et al., 2024) makes file-then-method-then-line localization explicit; RGFL (Sepidband et al., 2026) reasons to localize the fault region and anticipates the right-file-wrong-lines mode we observe; Sepidband et al. (2026, line-context) show that how the localized region is expanded changes repair; and Repoformer (2024) studies selective retrieval for repository-level code tasks. We concede this overlap directly: our own hypothesis registry (Appendix D, entry 4) records that RGFL anticipated the right-file-wrong-lines failure mode, so we claim it as prior art rather than as our finding. Anchor dose is, formally, thresholded line-level localization recall measured on the served pack, that is whether any gold pre-image line falls inside a packed chunk's span; we position it as a served-pack instance of a known FL quantity, not as a new metric, and our contribution here is the causal link from the packing flag to this quantity to resolution, not the quantity itself.

### 3.3 DeepDiscovery (He et al., 2026): a surface-opposite headline

He, Sun et al. ("From Fragments to Paths: Task-Level Context Recovery for Large Industrial Codebases," arXiv preprint, 22 Jun 2026) report, on the same benchmark, a headline that reads opposite to ours: recovering more task-level context raises SWE-bench Verified resolution from 70.4% to 78.6% (352/500 against 393/500, +8.2pp, paired McNemar $p<0.01$). We verified these numbers against the primary PDF. The two results are compatible because they occupy different budget regimes, and their own data shows it. Their budget is bounded but large (tens to hundreds of thousands of tokens); ours is bounded and small (12 slots). Their Table I reports micro-precision flat at about 10% across every compared method at 84 to 92% file-recall, so every system in their regime retrieves roughly ten times the gold set, which is precisely the condition our result says fails when the budget is tight and zero-sum. Their gains come from implementation-path completion via anchors plus multi-relational expansion, and they reject the reading that the gain is from exposing more content, which is closer to our depth arm than to shallow breadth. We read the pair as two sides of a budget-dependent regime boundary. Within the deployable slot-budget regime we sweep, the depth advantage grows with K rather than shrinking (on the clean strong-open Qwen3.6-27B, +4.8pp at K=4 to +9.2pp at K=40, within-model difference-in-differences $p=0.0003$; and independently on weak-open A3B, +2.0/+4.2/+6.2pp at K=4/12/40; Section 5.9), so more slots make within-file depth more, not less, advantageous. DeepDiscovery's breadth-recovery operates at budgets orders of magnitude larger (tens of thousands of tokens, far beyond K=40); any crossover to a breadth advantage is therefore necessarily non-monotonic and lies well outside the regime we test, and we neither observe nor claim it.

### 3.4 RAG-fusion deployment and retrieval-depth manipulations

Medrano et al. (2026) report an industry deployment of multi-signal RAG fusion tuned against IR metrics. Our result is not an evaluation of fusion retrieval quality (fusion against dense against BM25 as ranking systems) but a within-fusion ablation of a packing policy at fixed budget: the ranking source is held constant across arms and only the chunk-to-slot allocation changes. Their practice of tuning fusion configurations against IR metrics is an instance of the practice whose failure mode we measure. Meng et al. (When More Retrieval Hurts, 2025) show that increasing retrieved context degrades retrieval-augmented code review. Our manipulated variable is not the amount of retrieved context: both arms fill the same 12 slots, the depth arm is only slightly larger in tokens (about 5 to 6%, Section 6.2), and the effect survives in the stratum where the depth pack is the smaller of the two, so a retrieved-amount explanation is ruled out. Our knob is a recall-justified production default, not an experimenter-constructed retrieval-depth dial. We summarize these distinctions in the differentiation table below.

| work | intervention | retrieval unit | budget | dedup policy studied | task | paired within-instance |
|---|---|---|---|---|---|---|
| this paper | toggle file-dedup flag | file-keyed chunks | 12 slots (tight) | yes (the manipulation) | issue resolution | yes |
| DeepDiscovery (He et al., 2026) | add task-level context recovery | path/graph expansion | large (agent window) | no | issue resolution | yes |
| RAG-fusion deployment (Medrano et al., 2026) | tune fusion config | fused chunks | deployment | no | retrieval quality | no |
| When More Retrieval Hurts (Meng et al., 2025) | increase retrieval depth | passages | variable | no | code review | not reported |
| ContextBench (Li et al., 2026) | compare systems | mixed | mixed | correlational | context retrieval | no |

### 3.5 ContextBench (Li et al., 2026): the correlational sibling

ContextBench reports, across heterogeneous systems, that balanced retrieval wins over recall-maximizing retrieval, the correlational sibling of our finding. Correlational cross-system comparisons cannot separate packing policy from retriever quality, harness, or model; our single-flag within-system toggle can, and lands the same direction. We position our experiment as the controlled causal confirmation of ContextBench's correlational finding, not as an independent discovery.

### 3.6 Adjacent levers and effects

Pure length effects (Du et al., 2025) describe a monotonic tax as context grows with inert padding. Ours is not a length manipulation: both arms fill the same 12 slots, the 5 to 6% token gap favoring depth is controlled by stratification (Section 6.2), and the competing content is real retriever-ranked code rather than padding. Raju et al. (2026) find that long-context reasoning frequently degrades automated bug fixing, including for Qwen3-Coder-30B, one of our models; this varies the amount of retrieved context rather than the fixed-budget breadth-versus-depth allocation we hold constant. ArXiv 2607.09691 asks what context a coding agent needs and separates locating from acting; it manipulates code representation rather than the one-chunk-per-file packing policy at a fixed slot budget. Context compression (SWEzze, Jia et al., 2026) is an orthogonal lever (a trained compressor); ours is a free configuration flag. Line-context expansion (Sepidband et al., 2026) grows each snippet in place; ours reallocates a fixed inter-file slot budget.

### 3.7 Budgeted selection and granularity, and a naming disambiguation

Galimzyanov et al. (2025) study task-aware code-RAG retrieval design choices (chunking granularity, similarity scoring, split policy) under explicit compute budgets, the closest budgeted-retrieval design study to ours, but they do not manipulate file-level deduplication, so the specific packing-policy question we ask is not addressed there. On disambiguation: "deduplication" in Schelpe (2026) and the training-data-dedup literature means removing byte-identical or near-identical content, whereas our flag deduplicates by file identity within a ranked result list. Small-to-big, parent-document, and auto-merging retrieval (Dense X Retrieval, 2023) trade chunk granularity for context, and budgeted max-coverage and submodular passage selection do so under explicit token or top-k budgets, so these methods are applied under budgets. Their difference from our setting is narrower than "unbounded": they do not force the one-per-file zero-sum trade at 12 slots that produces our inversion. Dedup-OFF is best read as a crude within-file aggregation heuristic; a budget-matched small-to-big or parent-expansion baseline is future work. The question of how to aggregate multiple passages of one document for a downstream reader is long-standing in IR (passage-to-document score aggregation such as MaxP, Dai and Callan 2019, and PARADE, Li et al. 2020) and in end-to-end retriever-reader systems that learn to combine passages rather than toggling a packing flag (RAG, Lewis et al. 2020; Fusion-in-Decoder, Izacard and Grave 2021). Our contribution is orthogonal to these: we do not propose an aggregation method, but measure the downstream cost of one deployed file-keyed diversification default at a fixed budget.

## 4. Method

### 4.1 Task and grading

All experiments run on SWE-bench Verified, graded by the official SWE-bench Docker harness (`--cache_level instance`). Empty patches are counted as failures over the full n; the official harness drops them from its "submitted" denominator and we do not. Per-arm attempt (non-empty-patch) rates are reported alongside resolve rates, and Δ(resolve) is decomposed correctly in Section 5.2. A per-instance inclusion table with mutually exclusive statuses (resolved, unresolved, empty, harness-error, never-predicted) is released with the artifact so that every denominator in the paper is derivable from it (Section 4.6).

### 4.2 Retrieval stack and the manipulated flag

The retriever is a multi-signal fusion backend (`ragd`): dense embeddings (Qwen3-Embedding-8B, served via llama.cpp), a lexical signal (`lex_mode=ids`), and a graph signal (`w_graph=1.0`), fused into one ranking. The manipulated variable is one flag (`RAGD_QUERY_DEDUP_FILE`). ON, the production default, post-filters the ranking to at most one chunk per file before the top-12 cut; OFF serves the raw ranking. A per-request dedup override is plumbed client to wire to app to backend, so both arms are served by the same code path and index. The embedder, ranking weights, chunking, and slot budget are identical between arms. The chunker's size, overlap, and boundary policy and the chunk-line-count distribution are specified in the released artifact; a chunk-size-by-dedup sensitivity arm is future work.

The resulting pack shapes are dedup-ON at about 12 distinct files with one chunk each and dedup-OFF at about 5 files with multiple chunks per file. Dedup-ON is the file-level recall@k maximizer both in the index eval (0.666 to 0.817) and on the served packs (0.878 against 0.806 gold-file presence, Section 5.4).

### 4.3 Primary harness: single-shot, no tools

The primary harness is a single-shot runner (`local_dedup_ab.py` in the artifact). It issues one API call per instance (OpenRouter `/v1/chat/completions` for frontier models; local llama.cpp `llama-server` for open-weight models): the model receives the issue plus the 12-slot pack and must emit SEARCH/REPLACE edit blocks in one completion. Edits are applied to a pristine per-arm isolated checkout and captured as a git diff. There are no tools, no turns, and no reading beyond the pack. Packs are pre-fetched (500/500 instances, both arms), separating retrieval from generation. Decoding parameters were held identical across the two arms of each model: temperature 0.2 and a completion budget of `max_tokens=6000` for the non-reasoning arms (gpt-5.6-sol, Qwen3-Coder-30B, DENSE-1), with the reasoning-arm settings noted below; the local Qwen3-Coder-30B ran on a pinned llama.cpp `llama-server` (a non-reasoning coder model, no reasoning trace emitted) and the frontier arms on OpenRouter `/v1/chat/completions`. Decoding is sampled (temperature 0.2), not greedy, so the per-instance outcome carries sampling noise; because the paired McNemar test compares the two arms on the same instance with identical decoding and prefetched identical-source packs, this noise is common to both arms and does not correlate with the arm label. The exact slugs, providers, seeds where available, and the full prompt and applier are released with the artifact (Section 7, Appendix A). Identical pack rendering order was used across arms.

Reasoning-model handling was disclosed and fixed mid-program. Two harness defects initially produced 48 to 90% empty patches for reasoning-heavy models: the harness read the field `reasoning_content` where OpenRouter returns `reasoning`, and `max_tokens=6000` was a total budget spent on thinking before content. We fixed this by reading the correct field, setting `--max-tokens 16000` with `--reason-cap 2000` (a cap on requested reasoning tokens, honored by some providers and ignored by others), and persisting raw responses as an audit trail. Post-fix, DeepSeek-v4 empties fell to about 33% overall; measured per arm on the graded run they are 45.9% dedup-ON against 39.7% dedup-OFF (the asymmetry that defines its producibility channel, Section 5.2). The two DeepSeek figures cited in an earlier draft (an aggregate "38%/33%" and the per-arm 45.9%/39.7%) refer to the aggregate and the paired per-arm rates respectively; we report the paired per-arm pair throughout. GLM-5.2 and Kimi-k2.7-code remain incompatible with the regimen (the provider ignores the reasoning cap, or reasoning cannot be disabled), and are documented as reasoning models incompatible with the constrained single-shot regime, not as failed replications. We therefore define the target model population as models that can emit a SEARCH/REPLACE patch under a bounded single-shot completion, and present the incompatibilities as a scope limit rather than as neutral replications; the compatible inventory spans the weak-open Qwen3-Coder-30B-A3B, the strong-open Qwen3.6-27B, the frontier-closed gpt-5.6-sol, and DeepSeek-v4.

For Qwen3.6-27B (a reasoning model, reasoning held ON to match the gpt-5.6-sol tier) the completion budget was raised from `max_tokens=16000` to `32000` under a pre-registered, timestamped amendment (QWEN36-PREREG.md, Amendment 1, 2026-07-17) after 18/71 first-arm generations hit a bimodal runaway-reasoning truncation tail; the raise was sized blind to the ON/OFF resolution contrast, all 71 pre-amendment generations were discarded and both arms regenerated from scratch, and residual `fin=length` at 32k was pre-declared a capability failure counted as empty under the empties-as-failure primary. The verbatim amendment is in Appendix B.

### 4.4 Secondary harness: agentic CLI (boundary condition only)

The boundary condition of Section 5.5 uses an agentic harness: the `claude` CLI with the pack pre-injected via a pack-oriented system prompt, Read/Edit/Write enabled but no search tools, at most 10 turns, and unrestricted Read. Because Read is available and the agent iterates, the 12-slot budget no longer binds; this is a no-search agentic regime, not the single-shot regime, and its absolute rates are not comparable to Section 5.1. It is used only to locate where the effect stops. One arm produced with this harness was retired for a model confound and is reported only as a killed hypothesis (Sections 5.5, Appendix D); the lesson adopted from it is a methods control, namely to assert a per-transcript model census before grading any paired arms.

### 4.5 Statistics and reporting discipline

Paired test. McNemar's exact test on per-instance resolve outcomes; discordant counts are reported wherever a p-value appears.

Repository-clustered inference. Because SWE-bench Verified draws instances from a small set of repositories, instance-level exact tests can understate variance. For the primary arms we report, in addition to McNemar, a repository cluster bootstrap of the marginal Δ, leave-one-repository-out deltas, and per-repository discordant counts (Section 5.1). The primary gpt result survives clustering; the DENSE-1 dense control weakens (Section 5.3).

Primary estimand. The primary estimand, fixed in the committed E1–E2 allocation design (2026-07-14) and carried unchanged into every subsequent committed pre-registration, is the raw paired McNemar (empties as task failure); the conditional (both arms non-empty) is reported alongside it as a mechanism decomposition, because patch production is a post-treatment mediator. For gpt, Qwen, and DENSE-1 the effect survives both estimands, so their conclusion is estimand-invariant. DeepSeek is significant on the raw estimand but not on the conditional ($p=0.49$); we therefore present DeepSeek as supportive, producibility-channel evidence rather than as a co-equal confirmatory replication, and we do not label the family "confirmatory" without this caveat.

Multiplicity. The confirmatory single-shot family is gpt-5.6-sol, Qwen3.6-27B (the pre-registered open-weights replication), and DENSE-1, with DeepSeek supportive; the pre-registered multilingual pooled test (Section 5.10), though on a different benchmark, is folded into the same multiplicity family for completeness, for five primary raw p-values in all. Holm-Bonferroni across the five leaves the four SWE-bench Verified arms significant at $\alpha=0.05$ (gpt adjusted $p=0.0015$, DeepSeek 0.0016, DENSE-1 0.029, Qwen3.6-27B 0.029) and the multilingual pooled test non-significant (adjusted 0.056), exactly as it is uncorrected, so including it changes no conclusion. The budget×dedup dose-response (Section 5.9, both the clean Qwen3.6-27B K=4/K=40 axis and the A3B sweep) is reported as exploratory/secondary and is excluded from the confirmatory family. Secondary analyses (token-stratum, fix-locality, empty-asymmetry, interaction) are reported as exploratory and directional, not as additional confirmatory tests. Permutation tests use the stated statistic (marginal Δ), resample the exchangeable unit named in each case (instances within repository for cluster tests, arm labels within instance for the interaction), and use 5000 to 10000 replicates.

Underpowered nulls are reported as inconclusive with a minimum detectable effect (MDE) at 80% power ($\alpha=0.05$, exact McNemar on the observed discordant count), never as evidence of absence. Wilson CIs are used on small-n proportions.

### 4.6 Exclusions

Each paired comparison is computed over instances attempted in both arms (a gradable outcome, an empty patch, or a harness error); empty patches and harness errors are retained as failures, while instances never predicted in an arm are excluded pairwise because they cannot enter a paired test. This yields paired n = 500 (gpt-5.6-sol, and the weak-open A3B budget-sweep arms of Section 5.9), 499 (Qwen3.6-27B: one instance unmatched across arms; and DeepSeek-v4: one instance not predicted in dedup-ON), and 494 (DENSE-1: three never predicted and three graded incomplete, symmetric across arms). The released inclusion table gives per-arm, per-model error and exclusion counts and a sensitivity re-run excluding harness errors, so the reader can verify the exclusions are not asymmetric or post-treatment. Resolve rates are reported over the same paired n. Boundary-condition agentic comparisons are reported at their own n.

### 4.7 Multilingual generality harness and gold-validity gating (SWE-PolyBench)

The multilingual generality test (Section 5.10) runs the same single-shot dedup A/B on SWE-PolyBench (Rashid et al., 2025), using its task- and repository-stratified 500-instance subsample (SWE-PolyBench500), an execution-graded benchmark spanning Java, JavaScript, Python, and TypeScript (125 instances each). To keep the comparison honest and independent of us, we pre-registered (GENERALITY-BENCH-PREREG) the governing rules before generating any patch: the primary endpoint is the pooled two-model paired test with its N fixed in advance (G-2); generality is *supported* only if that pooled test is both significant and positive (G-3), otherwise we report an honest boundary (G-4); the grader is the benchmark's own published harness used without modifying its source, so host configuration (grading concurrency, memory) may be repaired but the grader's scoring, fail-to-pass parsers, and timeout constants may not (G-1); and no instance enters the A/B until it clears a gold-validity gate (G-5), described next.

Gold-validity gate. An execution benchmark is usable only where its own gold patch actually flips its own fail-to-pass tests under our frozen host configuration; an instance whose shipped gold patch does not pass is ungradeable and cannot inform a paired contrast in either direction. We therefore ran a gold-smoke pass over the non-empty union of attempted instances before any A/B grading. The first pass flipped 88.1% (267/303), below the pre-registered 95% floor. Per-repository diagnosis attributed the shortfall not to our method but to the shipped benchmark and its dependencies: a JavaScript grader whose Jest fail-to-pass parser does not credit passing tests (tailwindcss, a false negative), a dead model-download CDN (transformers), and out-of-memory or flaky Maven builds (rocketmq, svelte). Under G-1 we repaired only host configuration (reducing grading concurrency), never grader source, then re-smoked under the frozen configuration with one uniform retry. Recovery of the flaky cases left 280 gold-valid survivors against 23 exclusions (92.4% validity) in three cause classes: external-dependency drift, a shipped grader-parser defect, and genuine oracle or environment failures. This 92.4% remained below the pre-registered 95% floor; rather than chase the floor by relaxing the grader (which G-1 forbids), we froze and committed the full exclusion list with its cause classes before computing any A/B resolve and proceeded with the unmet floor recorded as a disclosed deviation, not a silent pass. The frontier gpt-5.6-sol arm reached 103 further instances the weak-open model had left empty; an incremental gold-smoke validated 62 and excluded 41 (the same cause classes plus image-build failures), extending the survivor set to 342. The full exclusion list and its cause classes were frozen and committed before any A/B resolve was computed, as an integrity anchor against post-hoc exclusion. Empty patches count as failures over the survivor set; the two TypeScript cells fall below an 80% patch-producibility floor and are reported descriptively. Five instances whose grading containers hung at the infrastructure level during the paired run were excluded from pairing as infrastructure artifacts, with a sensitivity analysis (specified when the hangs were found, not in advance) that instead counts them as failures (Section 5.10). The gate-failure findings themselves (a fail-to-pass parser that silently discredits passing tests, a dead dependency CDN, and self-inconsistent gold oracles) are reusable field warnings for anyone execution-grading multilingual SWE suites, and are the reason a gold-validity gate belongs in the method rather than in a footnote.

---

## 5. Results

### 5.1 Primary: single-shot cross-model table, with clustered inference

The primary evidence is single-shot, no tools, one shared harness, official Docker grading, empties as failures. The two packs are the treatment and differ by construction; everything else (retriever, ranking, embedder, budget, prompt, applier, decoding) is held identical between arms.

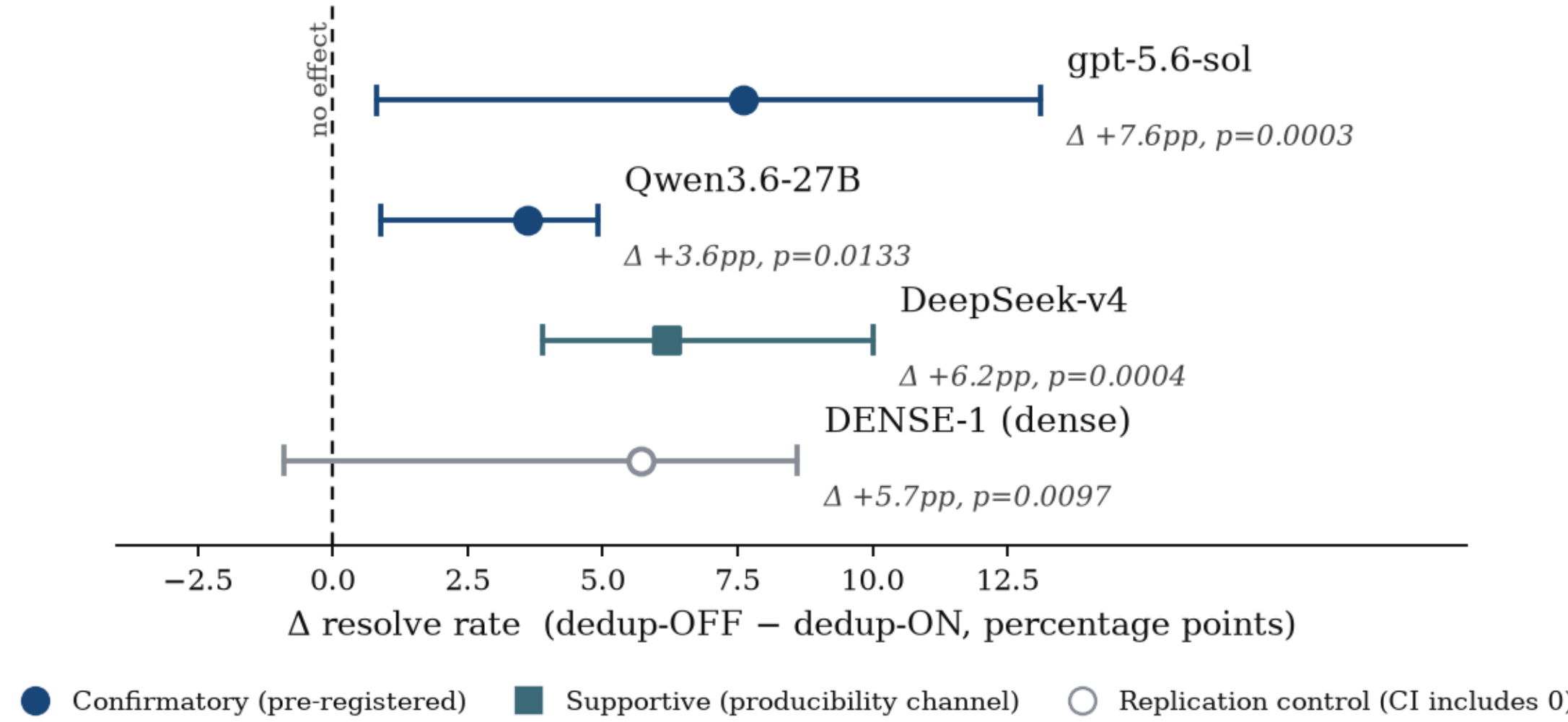

**Figure 2.** Single-shot Δ(resolve) (dedup-OFF minus dedup-ON, percentage points) with McNemar exact p-values and repository-cluster-bootstrap 95% confidence intervals, arms marked by status. Confirmatory: gpt-5.6-sol +7.6pp (p=0.0003, clustered CI [+0.8, +13.1]), Qwen3.6-27B +3.6pp (p=0.0133, clustered CI [+0.9, +4.9], pre-registered open-weights replication). Supportive (producibility channel): DeepSeek-v4 +6.2pp (p=0.0004, clustered CI [+3.9, +10.0]). Replication control (post-primary, dense): DENSE-1 +5.7pp (p=0.0097, clustered CI [−0.9, +8.6], includes zero). The vertical line marks no effect.

We report the raw paired McNemar as primary and the conditional (both arms non-empty) alongside it. Raw is primary because patch production is a post-treatment mediator: conditioning on "both non-empty" conditions on one channel the treatment acts through.

| model | vendor | pack | dedup-ON | dedup-OFF | Δ | disc (OFF/ON) | raw p | cond p | n |
|---|---|---|---|---|---|---|---|---|---|
| gpt-5.6-sol | OpenAI | fusion | 39.2% (196) | 46.8% (234) | +7.6pp | 73 / 35 | 0.0003 | 0.002 | 500 |
| DeepSeek-v4-pro | DeepSeek | fusion | 13.0% (65) | 19.2% (96) | +6.2pp | 53 / 22 | 0.0004 | 0.49† | 499 |
| Qwen3.6-27B (open weights) | Alibaba | fusion | 9.2% (46) | 12.8% (64) | +3.6pp | 33 / 15 | 0.0133 | 0.035 | 499 |
| DENSE-1 (retriever control) | OpenAI | dense | 41.5% (205) | 47.2% (233) | +5.7pp | 69 / 41 | 0.0097 | 0.0225 | 494 |
| Qwen2.5-Coder-32B | Alibaba | dense | 9.0% | 11.0% | +2.0pp | — | 0.688 | — | 98‡ |
| GLM-5.2 | Zhipu | — | — | — | — | — | — | — | —§ |
| Kimi-k2.7-code | Moonshot | — | — | — | — | — | — | — | —§ |

‡ Qwen2.5-Coder-32B: 98 paired of 100 attempted; underpowered and inconclusive (MDE table, Section 6.2), not evidence of absence. § GLM-5.2 and Kimi-k2.7-code are incompatible with the single-shot no-tools regime (provider-side unbounded or mandatory reasoning under a bounded completion budget), counted as neither replications nor non-replications (Appendix C).

† DeepSeek estimand disclosure. DeepSeek is significant on the raw analysis but not on the conditional (N=196, p=0.49). Its empty-patch rate is asymmetric (dedup-ON 45.9% against dedup-OFF 39.7%, p=0.025), so the

treatment acts through patch producibility, a channel the conditional analysis removes by construction. We present DeepSeek as supportive producibility-channel evidence, not a co-equal confirmatory replication.

Qwen3.6-27B robustness (truncation). Both Qwen3.6-27B arms are empty-heavy under the constrained single-shot reasoning regime (58.1% ON / 57.9% OFF, symmetric), so the raw 9.2%/12.8% resolve rates sit atop a high empty floor. The complete-case analysis restricted to pairs where both arms produced a completion (n=269, dropping truncated pairs) gives +6.3pp (p=0.0095), and Manski bounds counting truncated-empties as failure versus success span +3.0 to +3.6pp, so the effect survives the worst case. Truncation is symmetric across arms (`fin=length` ON 25.8% / OFF 25.5%), and of dedup-OFF's 33 discordant wins only 5 are cases where dedup-ON truncated, so the advantage is a within-file-depth effect, not differential censoring. The complete-case +6.3pp is the robustness lead.

Repository-clustered inference (twelve repositories). We report a repository cluster bootstrap and leave-one-repository-out (LORO) for all three single-shot fusion arms and the dense control. The primary gpt effect survives clustering: 95% CI [+0.8, +13.1]pp (one-sided clustered P(Δ≤0)=0.018), LORO deltas +6.1 to +9.2pp, and no single repository drives it (dropping the two largest contributors, django and sympy, leaves +6 to +9pp). Qwen3.6-27B, the pre-registered open-weights replication, also survives repository-clustered inference: cluster-bootstrap 95% CI [+0.9, +4.9]pp (one-sided P(Δ≤0)=0.007), and all twelve leave-one-repository-out folds stay positive (+2.2 to +4.0pp, the smallest dropping django), so no single repository drives it. DeepSeek is robust to clustering (CI [+3.9, +10.0]pp). The DENSE-1 dense control weakens: its clustered 95% CI is [−0.9, +8.6]pp and includes zero, which we carry into the retriever-generality claim (Section 5.3). We note twelve clusters is a small number for a percentile cluster bootstrap, so these intervals are approximate; a small-cluster or wild-bootstrap treatment is a reasonable robustness extension. The effect replicates in direction across three model arms and both embedding retrievers, though Qwen3.6-27B and the A3B anchor share Alibaba/Qwen lineage and DeepSeek is a separate but adjacent open lineage, so the open-weights arms are not fully independent draws and gpt-5.6-sol is the one cross-family confirmatory point; it is not a universality claim, and local dense Qwen2.5 and two reasoning-only models are inconclusive or incompatible.

### 5.2 Attempt-rate decomposition (corrected)

Resolve is a product, not a sum: with A the attempt rate (one minus the empty rate) and Q the correct-given-attempt rate, R = A·Q, and the difference decomposes exactly as ΔR = Q_on·ΔA + A_on·ΔQ + ΔA·ΔQ. The earlier additive form ΔR = ΔA + ΔQ is incorrect. Per arm:

All deltas use the paper's OFF-minus-ON convention. Δattempt(OFF−ON) is the attempt-rate gain (attempt = one minus the empty rate); the three rightmost columns give each decomposition term as a signed percentage-point contribution to ΔR and, in parentheses, its share of ΔR.

| model | empty ON | empty OFF | Δattempt (OFF−ON) | ΔR (OFF−ON) | producibility Q_on·ΔA | quality A_on·ΔQ | interaction ΔA·ΔQ |
|---|---|---|---|---|---|---|---|
| gpt-5.6-sol | 14.0% | 13.2% | +0.8pp (n.s.) | +7.6pp | +0.4pp (5%) | +7.2pp (94%) | +0.1pp (1%) |
| DENSE-1 (dense) | 13.6% | 12.3% | +1.3pp (n.s.) | +5.7pp | +0.6pp (11%) | +5.0pp (88%) | +0.1pp (1%) |
| DeepSeek-v4 | 45.9% | 39.7% | +6.2pp (p=0.025) | +6.2pp | +1.5pp (24%) | +4.2pp (68%) | +0.5pp (8%) |

The quality channel (correct-given-attempt) carries 88 to 94% of the gain for gpt-5.6-sol and DENSE-1; a secondary producibility channel is material only for DeepSeek (about 24%), whose attempt-rate gain is the only one that is statistically significant. The weak-open A3B arm's attempt/quality split is reported in Section 5.9 (empty 41.6% ON / 38.8% OFF) rather than here, since its absolute levels are truncation-contaminated; its +4.2pp likewise runs mostly through quality, but we treat A3B as a Δ-trend result, not a levels result. We therefore do not claim the effect runs through "producibility for weaker models"; where it is cleanly measurable the gain is almost entirely quality. For DeepSeek the extra failures under breadth are real and asymmetric, and we verified them from logs: over about 40 producibility-failure cases, the majority are a SEARCH/REPLACE attempt whose anchor does not match the served context (20 of 38: text produced, no edit applied; only 2 fully empty). The unifying reading is a hypothesis, not a proven identity: models edit better what they are shown enough of.

### 5.3 Retriever scope: two embedding retrievers, and a lexical reversal

An earlier draft eliminated distractor-removal via an n=100 agentic mechanism table; a per-transcript model census retired that contrast (its two sub-arms ran different models, the confound of Section 5.5), so we withdraw that elimination and treat distractor-removal as a live candidate mechanism entangled with anchor dose by the knob's construction (Section 5.4). We do not claim the controls eliminate the distractor account.

The at-power retriever-generality control is single-shot DENSE-1: gpt-5.6-sol on dense-only packs, n=494, dedup-ON 41.5% against dedup-OFF 47.2%, Δ +5.7pp, discordant 69 against 41, raw p=0.0097, conditional p=0.0225. We label DENSE-1 a replication and generality control, not a pre-registered confirmatory arm: its harness and analysis design were committed after the primary gpt effect had been observed (about a day later), though before DENSE-1's own run was executed, so it was specified before its own data but not registered before the primary result. A single-signal dense retriever reproduces the effect at the instance level, but its advantage does not survive repository-clustered inference (cluster-bootstrap CI [−0.9, +8.6]pp includes zero). The fusion-versus-dense interaction is not detected (permutation p=0.39), which is a fail-to-reject on two same-embedder retrievers, not an equivalence result. We therefore read DENSE-1 as a directional replication whose evidentiary weight is limited by both its post-primary design and its non-significant clustered interval.

We tested generality across retrieval paradigms in two stages. An exploratory n=100 BM25 pilot (gpt-5.6-sol, keyword-mode ranking) was run first, and its result (dedup-ON 76% against dedup-OFF 74%, Δ −2pp, p=0.69, already the reversal direction, though uninformative at that n) was committed to the repository on 2026-07-14 together with a direction rule for the retriever-agnosticism test (Appendix B). We then ran the powered n=500 version of the same A/B with a repository-clustered analysis, an equivalence (TOST) test at a 5pp margin, and a commitment to report a null or reversal as a bound; this analysis plan was fixed in a dated design note before the powered run but was not committed before it (a Class-2 note, Appendix B), so we describe the powered arm as a powered confirmation of a committed pilot direction, not as pre-registered. The powered result confirms the pilot's reversal:

| retriever (gpt-5.6-sol) | dedup-ON | dedup-OFF | Δ(OFF−ON) | cluster-bootstrap 95% CI |
|---|---|---|---|---|
| fusion (headline) | 39.2% | 46.8% | +7.6pp | survives (Section 5.1) |
| dense (DENSE-1) | 41.5% | 47.2% | +5.7pp | [−0.9, +8.6]pp |
| BM25 (lexical) | 37.2% | 34.0% | −3.2pp | [−5.8, −0.3]pp |

BM25 reverses the sign. The evidence that BM25 differs from the embedding retrievers is the cross-retriever interaction, which is significant: fusion minus BM25 is +10.8pp (95% CI [+5.3, +16.4]pp) and dense minus BM25 is +8.9pp (95% CI [+3.2, +14.5]pp), both excluding zero. The equivalence (TOST) test at a 5pp margin (the margin fixed in a Class-2 design note, Appendix B) does not support equivalence between BM25 and either embedding retriever, but we lean on the significant interaction rather than on the failed equivalence test, since failing to establish equivalence is not itself evidence of a difference. This BM25 reversal is not a recall-floor artifact: BM25's served-pack gold-file recall is lower (0.69 dedup-ON against 0.61 dedup-OFF) than the embedding retrievers, but BM25 still resolves at 34 to 37%, comparable to fusion, so it is a functional retriever. The mechanism corollary is consistent with the reversal (Section 5.4): BM25's depth arm raises within-file gold-line coverage only +0.029 (0.059 to 0.088), against fusion's +0.089, so the anchor-dose gain that the effect tracks barely materializes under BM25. We therefore state, conservatively, that the effect holds for the two embedding retrievers tested and does not extend to this lexical BM25 configuration, where it reverses. We do not claim a clean embedding-versus-lexical paradigm boundary: this rests on one BM25 configuration on one model, fusion itself mixes lexical and graph signals with dense, and the DENSE-1 dense arm's own clustered interval includes zero. Generality beyond the two embedding retrievers tested is not established. What the BM25 arm does establish, as a two-stage result whose reversal direction was committed at pilot scale before the powered run under a committed direction rule designed to be able to refute cross-paradigm generality, is that the effect is not a universal document-count law: at least one functional retriever shows the opposite sign, and its anchor-dose gain is correspondingly small.

### 5.4 Mechanism: anchor dose (an oracle diagnostic and partial mediator) and a selection control

Coverage accounting was recomputed on the served packs (n=500; gold lines from each instance's pre-image gold hunks). The depth arm roughly doubles gold-line coverage (fusion 0.090 to 0.179, Δ about +0.089; dense 0.083 to 0.187) while finding the gold file slightly less often (present in 0.878 against 0.806 of fusion packs; 0.890 against 0.802 dense). These served-pack figures supersede an earlier live-re-fetch computation (0.130/0.235 coverage; 0.92/0.98 file-find) that did not reproduce against the packs the models saw. The coverage direction and near-doubling are unchanged, but the file-find ordering was corrected: the earlier numbers had dedup-OFF finding the gold file more often (0.92 against 0.98), whereas the served-pack recount has dedup-ON finding it more often (0.878 against 0.806), which is the direction the recall-maximizer premise predicts.

Anchor dose is an oracle diagnostic: a visible gold anchor requires the gold pre-image lines, which are known only post-patch, so anchor dose is a retrospective measurement, not an inference-time signal, and it is line-level localization recall measured on the served pack (Section 3.2). We report it as a supported but partial mediator, and the mediator pattern holds across all four arms. Resolution is far higher when an anchor is visible than when it is absent (gpt 54.3% against 35.2% dedup-ON, 61.2% against 38.8% dedup-OFF; dsv4 30.5% against 8.4%, 38.8% against 8.4%; Qwen 16.2% against 4.1%, 29.8% against 2.8%; DENSE-1 50.5% against 39.2%, 57.5% against 40.1%). The depth arm shifts about 15 percentage points of instances into the anchor-visible stratum (fusion 21.0% to 35.6%, dense 19.8% to 38.6%), which is the dose. A within-stratum residual remains (dedup-OFF beats dedup-ON among anchor-visible instances by +7 to +13.6pp, largest for Qwen), so anchor dose is a substantial but partial mediator, descriptively about 40% of the effect, and the residual is unexplained. We stop short of a full causal claim: the knob is compound, and anchor dose is not decomposed against distractor-removal; a splice or placebo intervention that holds file count fixed and varies anchor dose is the identifying test, and an inference-time surrogate for anchor dose is future work. We do not claim the mediator is "chiefly upstream" or arm-invariant.

Selection versus depth (a designed control). Because the flag also determines which within-file chunk is kept (the argmax), the harm could be that dedup-ON keeps the wrong chunk (a signature or interface chunk rather than the gold body) rather than that it serves fewer chunks per file. We ran a random-chunk-per-file control on Qwen3-Coder-30B (n=500, three seeds). Its decision rule was fixed in a dated design note before any generation, but the note was not committed to the repository before the run, so we do not call this control pre-registered (Appendix B); its three pre-specified seeds were all run and reported (seed-averaged, below), with no variant discarded. Arm B holds the dedup-ON file set fixed and, for each file, serves a uniformly random non-argmax chunk from that file's candidate pool instead of the argmax. If the harm were bad selection, random reselection would recover toward the depth arm; if the harm is depth, random reselection would not help. The outcome is unambiguous:

| Qwen3-Coder-30B arm | resolve | contrast |
|---|---|---|
| A: dedup-ON (argmax, existing) | 6.6% | — |
| B: random-chunk-per-file (new, seed-averaged) | 3.0% | B − A = −3.6pp (per-seed McNemar $p \le 0.017$; cluster-bootstrap CI [−5.3, −0.3]pp) |
| C: dedup-OFF (depth, existing) | 12.4% | B − C = −9.4pp ($p < 1e\text{-}4$) |

(Arms A and C reuse this random-chunk control's own Qwen3-Coder-30B dedup-ON/OFF runs; their resolve rates, 6.6% and 12.4%, differ slightly from the same model and config in the Section 5.9 K=12 budget-sweep run, 7.4% and 11.6%, as independent samples at temperature 0.2, and the depth-over-breadth direction is identical in both.) Random reselection is significantly worse than the argmax it replaces, not better. Argmax within-file selection is therefore beneficial, which refutes the argmax-selection-bias alternative that dedup-ON underperforms because it keeps the wrong (signature or interface) chunk. What the control does not do is identify within-file depth as the cause of the A-versus-C gap: Arm B holds the dedup-ON file set fixed, whereas dedup-OFF (Arm C) serves a different, smaller file set (about five files against twelve) with different rank, score, and same-file adjacency, so the control speaks to chunk selection within a fixed file set, not to the file-count change that separates A from C. Separating within-file depth from the file-count change, and from a same-file-contiguity effect in which adjacent chunks reconstruct a coherent span, requires a file-set-fixed depth control and an interleave or merge control that we did not run; these are the remaining identification steps and are future work. Arm B was verified sound before inference: its packs are well-formed and slightly smaller in tokens than Arm A, it used the identical harness and model, and its elevated empty rate decomposes into truncation, no-op edits, and unanchorable SEARCH blocks. Two further caveats: Arm B was run only on Qwen3-Coder-30B, the weakest and highest-empty-rate model, and its gap runs substantially through producibility with an underpowered conditional (p=0.21). It refutes argmax-selection bias; it does not establish depth over file count.

A gold-file-presence stratification of the primary arm's discordants is, however, consistent with a within-file-depth reading, on the subset where file membership is held fixed by nature. Classifying the gpt discordant pairs by whether the gold file is present in each arm's served pack, 82% of dedup-OFF's wins (60 of 73) fall in the stratum where the gold file is present in both arms' packs (resolve 44.5% against 52.0%, +7.5pp), so in the majority of wins dedup-OFF is not newly finding the gold file but covering an already-present file more deeply. Only 7 of 73 wins occur where the gold file is in neither pack (a distractor-removal residual), and none where dedup-OFF alone adds the gold file. This is observational and conditions on a partly treatment-affected variable (pack membership), so it is supportive rather than an identification, but it points the same way as the depth reading and locates a distractor-removal residual consistent with the mediator being partial.

### 5.5 Boundary condition: the effect is not detected under unrestricted Read

The recall trap is a property of pack-as-sole-context deployments. When the same packs are handed to an agentic harness with unrestricted Read (Read/Edit/Write, at most 10 turns), the fixed budget stops binding and we no longer detect the effect. Every model-matched agentic dedup comparison is a null, including the arm on the headline retriever, fusion, run with sonnet-5 in both sub-arms: n=499, dedup-ON 65.9% against dedup-OFF 64.5%, Δ −1.4pp, discordant 28/35, McNemar exact p=0.45. This is a powered null (MDE 4.5pp, below the +7.6pp single-shot fusion effect); its point estimate is slightly negative, so the precise statement is that the effect is not detected and is bounded below its single-shot size, not that a reversal is established. We note that this boundary is observed under sonnet-5 and opus-4-8 agentic harnesses, not under the single-shot models, so it is a boundary on the harness with a model change confounded in; a same-model single-shot-versus-agentic factorial is future work. The dense and BM25 agentic arms agree (sonnet-5 dense n=499 −1.8pp p=0.23; opus-4-8 dense and BM25 n=100 −2pp each, both underpowered). Tool-using code assistants that query retrieval in a loop and can open any file fall in this free-reading regime, where our positive finding does not apply.

Retired arm (Appendix D). An earlier draft reported an agentic fusion result of +18.2pp (n=499) as a robustness observation. A per-transcript model census showed its two sub-arms ran different models (dedup-ON sonnet-5, dedup-OFF opus-4-8) from a launch-script default mismatch: it measured opus against sonnet, not dedup. The matched-model replacement is the powered null above. We report the +18.2pp only as a killed hypothesis and a methods note.

### 5.6 Concentration sweep (withdrawn, confounded)

An earlier draft presented an inverted-U concentration sweep on the n=100 slice. It is withdrawn: the sweep ran on the agentic harness and mixed the two confounded models, and within a single model the served points are flat with no clean inverted-U. We make no numeric-optimum or inverted-U claim, and the anti-length argument in Section 3.6 rests on the token-stratum control of Section 6.2 alone.

### 5.7 Failure taxonomy and the parse-apply-test funnel

We classify each single-shot gpt-5.6-sol dedup-ON instance by what its pack contained relative to the gold hunks (n=497 graded, 301 failures; Wilson 95% CIs on the share of failures).

| pack-content class | n | resolve rate | share of failures | class prevalence |
|---|---|---|---|---|
| A — gold file absent from pack | 60 | 15.0% | 16.9% [13.1, 21.6] | 12.1% |
| B — gold file present, gold lines not in any chunk | 359 | 37.3% | 74.8% [69.6, 79.3] | 72.2% |
| C — gold lines fully covered (failed anyway) | 78 | 67.9% | 8.3% [5.7, 12.0] | 15.7% |

Class B's 74.8% share of failures is close to its 72.2% prevalence, so it is not disproportionately enriched; we therefore do not describe failures as "dominated by an allocation problem." The load-bearing evidence is the monotone dose-response: resolve rate climbs 15.0% to 37.3% to 67.9% across the classes, and DENSE-1 shows that supplying lines via the single-shot dense arm raises resolution. The right-file-wrong-lines mode (class B) is anticipated by RGFL (Sepidband et al., 2026), cited as prior art.

Parse-apply-test funnel (defusing the format artifact). Because the single-shot harness requires verbatim SEARCH anchors, the effect could in principle be an edit-format artifact rather than a repair-quality effect. From the raw logs of the arms where they are available we build the funnel generate → applied → test-pass and compare the depth and breadth arms among patches that actually applied. Depth still wins on test-pass conditional on applying: DeepSeek 23.9% against 31.9% (+8.0pp), DENSE-1 48.4% against 55.1% (+6.7pp), and gpt-5.6-sol 47.9% against 53.1% (+5.2pp; pre-registered logged replication, Appendix B). Among patches that parse and apply, the depth arm is more often correct, so the effect is not merely that depth produces more anchorable patches. The original gpt headline run's responses were not logged, so its own apply stage cannot be split; the gpt funnel row is computed from a pre-registered logged re-run of the identical design (same served packs, same prompt, same n) executed for this purpose. The replication run's own resolve contrast, whose publication regardless of value was committed before the run (Appendix B), was Δ = +6.4pp (67 against 35 discordant, exact p = 0.0020, repository-clustered CI [+4.9, +11.3]pp); under the one-run-per-curve rule it is reported here as a second sampling draw of the same estimand and does not replace the headline estimate, which remains the original run's +7.6pp. It meets the pre-registered directional criterion of Appendix B. One caveat remains: DeepSeek's format-and-apply loss is large (about 40%, its real producibility channel); a format-orthogonal replication (whole-file or unified-diff output) is future work.

### 5.8 Contamination-dosage stratification

Because SWE-bench Verified repositories are plausibly in pretraining data and the depth arm exposes more verbatim gold lines, the effect could be confounded with memorization. We stratified by repository popularity (GitHub stars, a coarse memorization-pressure proxy). The per-repository dedup effect is essentially uncorrelated with popularity (Spearman ρ between stars and effect ≈ +0.12 across repositories): the largest effects span the full popularity range (astropy at about 4k stars, +9pp; scikit-learn at about 60k, +10pp) while some high-star repositories are null. A median split shows only a mild positive tilt (high minus low popularity about +2.7pp averaged across arms), and the effect remains substantial and mostly significant in the low-popularity stratum (+3.4 to +6.4pp). There is thus no popularity dose-response, so contamination amplification does not appear to explain the effect. We report this as a check, not a proof: every SWE-bench Verified repository is a popular public Python project, so the proxy has restricted range and cannot exclude a floor of contamination shared across all repositories. A post-training-cutoff benchmark is the decisive control and is future work.

### 5.9 Budget×dedup dose-response: the trap grows with the slot budget

The clean dose-response axis is the strong-open Qwen3.6-27B run single-shot at two budgets, K=4 and K=40, on the same fusion packs and harness as the primary, under the 32k completion budget that removes the truncation floor (Section 4.3); unlike the exploratory weak-open sweep below, its levels are not truncation-contaminated. At K=4, dedup-ON resolves 43/499 against dedup-OFF 67/499 (Δ +4.80pp, McNemar exact p=1.2e-4); at K=40, 44 against 90 (Δ +9.20pp, p=1.7e-9). The depth advantage roughly doubles across the budget, and we test that growth directly rather than by eye: a within-model difference-in-differences (the K=40 Δ minus the K=4 Δ, bootstrapped over repository clusters) gives ΔΔ = +4.40pp, 95% CI [+2.65, +6.23], one-sided p=0.0003. "The trap grows with the slot budget" is therefore an inference, not a monotone eyeball, and it is on a clean model rather than the truncation-contaminated weak-open arm.

Per the program's one-run-per-curve discipline, this axis is the two clean K=4 and K=40 runs only. The Qwen3.6-27B K=12 point (+3.6pp, Section 5.1) is from an earlier, higher-empty regime (about 58% empty under a lower completion budget) and is deliberately not grafted onto the curve; its lower value is that regime's empty-floor attenuation, disclosed, not a non-monotonicity of the clean axis. Neither dose-response axis in this section carries a pre-registration: the clean K=4/K=40 runs were identified in advance as the confirming test by the exploratory weak-open sweep's monotone trend, but no design document was committed before them, so we report the growth result as a within-model replication of an exploratorily observed trend, outside the confirmatory family (Section 4.5, Appendix B).

An exploratory second dose-response on the weak-open Qwen3-Coder-30B-A3B at three budgets K∈{4,12,40} (n=500 each, same fusion packs and harness) shows the same monotone direction independently:

| K | dedup-ON | dedup-OFF | Δ(OFF–ON) | disc (OFF/ON) | McNemar exact p |
|---|---|---|---|---|---|
| 4 | 6.8% (34) | 8.8% (44) | +2.0pp | 26 / 16 | 0.164 (n.s.) |
| 12 | 7.4% (37) | 11.6% (58) | +4.2pp | 33 / 12 | 0.0025 |
| 40 | 6.8% (34) | 13.0% (65) | +6.2pp | 42 / 11 | 2.2e-05 |

dedup-ON is flat across K (6.8 to 7.4%) while dedup-OFF climbs (8.8 to 11.6 to 13.0%): extra slots spent on within-file depth keep paying off, extra slots spent on file breadth do not. We report this as a Δ-trend, not as absolute levels: A3B's empties are truncation-contaminated (its 6000-token completion cap, the same failure family the Qwen3.6-27B amendment fixes), so the levels are depressed and only the direction and the monotone growth in Δ are load-bearing. The K=4 null is inconclusive, not evidence of absence: at the observed discordant volume (42 pairs) the minimum detectable effect at 80% power is ≈3.6pp (exact McNemar, computed as for the Section 6.2 MDE table), above the observed +2.0pp (power at the observed split ≈32%). This weak-open sweep is exploratory and truncation-contaminated; the clean-model K=4/K=40 axis above is the confirming test it identified, and that test confirms the direction with a significant growth statistic (ΔΔ p=0.0003). A3B's attempt/quality split is reported here rather than in the Section 5.2 table: at K=12 its empty rate is 41.6% (ON) / 38.8% (OFF), so the +4.2pp gain runs partly through producibility and its absolute levels are truncation-contaminated.

### 5.10 Multilingual generality: a mapped boundary (SWE-PolyBench)

Under the pre-registered analysis (GENERALITY-BENCH-PREREG, endpoints fixed before data), multilingual generality is *not supported*. Pooled across two models and four languages on SWE-PolyBench, the trap-direction advantage is Δ = +2.59pp (39 dedup-OFF-only against 23 dedup-ON-only discordant wins, N=617 model×instance pairs), exact McNemar p = 0.056: directionally positive but below the pre-registered significance threshold, so the binding branch is G-4, an honest boundary. The test stacks two independent single-shot dedup A/Bs (the weak-open Qwen3-Coder-30B-A3B and the frontier gpt-5.6-sol) on SWE-PolyBench (Java, JavaScript, Python, TypeScript; 125 instances each), graded by the benchmark's own off-the-shelf execution harness, empties as failure, over gold-valid survivor instances (Section 4.7), and was pre-registered to be able to refute cross-language generality. It was designed for a minimum detectable effect of 2.8pp at 80% power on a planned pool of N≈1000 (500 instances × two models); the gold-validity gate (Section 4.7) reduced the realized pool to N=617, so the realized MDE at 80% power, on the observed discordant volume (62/617), is ≈3.6pp.

| language | A3B: n / Δpp / disc OFF–ON / p | gpt-5.6-sol: n / Δpp / disc OFF–ON / p |
|---|---|---|
| Java | 89 / +0.0 / 2–2 / 1.00 | 110 / −2.7 / 7–10 / 0.63 |
| JavaScript | 86 / +0.0 / 1–1 / 1.00 | 111 / +8.1 / 12–3 / 0.035 |
| Python | 80 / +10.0 / 8–0 / 0.008 | 79 / −1.3 / 6–7 / 1.00 |
| TypeScript† | 25 / +0.0 / 0–0 / — | 37 / +8.1 / 3–0 / 0.25 |
| pooled (model) | 280 / +2.9 / 11–3 / 0.057 | 337 / +2.4 / 28–20 / 0.31 |

† Per-cell p-values are exploratory and uncorrected. The two TypeScript cells had 0 and 3 discordant pairs respectively, below the pre-registered patch-producibility ratability floor (80%), and are reported descriptively.

Two of the eight cells reached nominal significance in the trap direction (Qwen3-Coder-A3B on Python, +10.0pp, uncorrected p=0.008; gpt-5.6-sol on JavaScript, +8.1pp, uncorrected p=0.035). These are exploratory, uncorrected subgroup contrasts: under Holm correction across the eight cells neither survives (smallest adjusted p = 0.062), and the two cells are disjoint in both model and language, a pattern as consistent with noise concentrated in small cells as with a heterogeneous real effect. We do not claim the effect is established in any individual language.

The two model pools, run independently, yielded similar pooled point estimates (+2.86pp and +2.37pp), neither individually significant. The combined estimate (+2.59pp) sits below the realized minimum detectable effect (≈3.6pp at 80% power on the observed 62/617 discordant pairs; the 2.8pp figure was the design target for the planned N≈1000 pool, before the gate exclusions), so as run the study is underpowered relative to plan and the observed estimate is smaller than even the realized MDE. Because the same PolyBench instance can appear as a pair under both models (276 of 341 instances do), the 617 pooled pairs are not fully independent across models; an instance-clustered bootstrap over the 341 instances leaves the verdict unchanged (95% CI [0.0, +5.3]pp, one-sided P(Δ≤0)=0.031, an interval that still reaches zero). The data are consistent with a small positive effect below our detection threshold; we do not claim one.

The most informative cell is the null. gpt-5.6-sol on SWE-PolyBench Python is flat (−1.3pp, p=1.00), while the same model on SWE-bench Verified (also Python) showed +7.6pp (p=0.0003). Language is therefore not the moderator. Whatever conditions the trap, it operates at the level of the instance pool and pack composition, not the surface language. This cell also blocks the tempting reading of A3B's Python cell as a same-language quasi-replication of the core SWE-bench result: gpt-Python shows that the inference does not go through.

Five instances whose grading containers hung at the infrastructure level during the paired run were excluded from pairing as infrastructure artifacts (Section 4.7); a sensitivity analysis (specified when the hangs were found, not in advance) that instead counts them as dedup-OFF failures gives +2.41pp, p=0.077, the same conclusion.

We therefore report multilingual behavior as a mapped boundary, not a confirmed generalization. The trap is established for SWE-bench Verified Python single-shot packs; across languages the pooled estimate is positive, small, and below our detection threshold, with heterogeneity that tracks the instance pool rather than the language. This gives the generality section the same epistemic shape as the rest of the paper: a strong scoped core with honestly mapped edges (BM25, agentic loops, larger budgets, and now multilingual pooling), rather than a generality that conveniently confirms everywhere.

---

## 6. Discussion

### 6.1 A registered and audited program and its retired hypotheses

This result is the surviving hypothesis of a registered and audited program: the major confirmatory arms carry pre-data design commits, Appendix B states each document's commit and downgrades the label wherever the ledger does not support it, and several sibling directions were retired by pre-specification, hostile prior-art hunts, or our own audits. We report the registry in Appendix D (hypothesis, pre-registration reference, data used, disposition type: refuted, underpowered, confounded, not-novel, or superseded). A ledger of dead hypotheses is a transparency and multiplicity record, not itself a scientific contribution, which is why it is an appendix rather than a headline claim. Two entries are our own audits retiring earlier internal claims: the agentic +18.2pp arm (a model confound, Section 5.5) and the fusion-specificity claim (DENSE-1 dense reproduces the direction, Section 5.3). The claim is carried by two confirmatory single-shot arms that survive both estimands and repository clustering (gpt-5.6-sol, and the pre-registered open-weights replication Qwen3.6-27B), with DeepSeek supportive (producibility channel) and DENSE-1 a directional dense replication whose clustered interval includes zero. Holm-Bonferroni across the four single-shot A/Bs leaves all four raw p-values significant; the surviving claim rests on the two clustering-robust confirmatory arms and does not depend on any single arm.

### 6.2 Threats to validity

Empty-patch / attempt-rate confound (addressed, Section 5.2). The corrected product decomposition shows the quality channel carries 68 to 94% of every arm's gain; only DeepSeek has a material producibility share.

Producibility versus quality (Section 5.7). The parse-apply-test funnel shows depth wins on test-pass conditional on applying for every arm with raw logs, so the effect is not reducible to an anchor-format artifact; the gpt apply stage, unlogged in the original run, is supplied by a pre-registered logged replication (+5.2pp on test-pass conditional on applied, Section 5.7), and a format-orthogonal replication is future work.

Pack-size (token) confound. "Fixed budget" is slot-fixed, not token-fixed. The depth arm is slightly larger: fusion dedup-ON 1451 against dedup-OFF 1525 tokens (+5.1% under o200k), dense 1322 against 1402 (+6.0%). Native-tokenizer recounts reproduce the gap (Qwen3 native +5.2%, DeepSeek-V3 native +5.6%), and the "dedup-OFF is smaller" subset the control relies on is 175 to 180 of 500 regardless of tokenizer, so tokenizer choice is immaterial. Restricted to the 179 instances whose dedup-OFF pack is smaller in tokens, the effect survives (gpt discordant 23/9, $p=0.020$; Qwen 18/4, $p=0.0043$), and discordance direction is uncorrelated with the token delta (MWU $p \approx 0.99$). Pack size is a reported covariate, not the driver. A separate objection, that slots are not an information budget and depth may add repeated rather than unique exposure, is addressed only in part here (we report per-slot token sizes); a re-render-to-equal-tokens control and pairwise chunk-overlap and unique-line counts are future work.

Fragmentation versus dedup. At about 120 tokens per slot, dedup-ON serves small single-fragment views and dedup-OFF reconstructs a larger contiguous view. Whether the lesson is "do not use tiny one-fragment-per-file packing" rather than "disable dedup" is not separated here; a chunk-size-by-dedup interaction and budget-matched parent-expansion baselines are future work.

Positional salience. Dedup changes chunk sequence, adjacency, and anchor slot positions (Lost-in-the-Middle). We report slot-wise anchor position distributions in the artifact; a position-randomized robustness arm is future work.

Contamination cue. Addressed by the popularity stratification (Section 5.8): no dose-response with popularity, effect persists in low-popularity repositories; a post-cutoff benchmark is the decisive control.

Construct: what the flag confounds. Toggling dedup changes file count, chunks-per-file, served rank/score, within-file selection, and verbatim-anchor availability at once (Section 2). The random-chunk control (Section 5.4) isolates the selection component and refutes it; a splice/placebo intervention to isolate anchor dose from distractor-removal is future work.

Statistical power of the controls. MDE at 80% power for every null:

| null arm | n | observed Δ | discordant (OFF/ON) | raw p | MDE @80% | powered? |
|---|---|---|---|---|---|---|
| agentic fusion (sonnet-5) | 499 | −1.4pp | 28 / 35 | 0.45 | 4.5pp | yes |
| agentic dense (sonnet-5) | 499 | −1.8pp | 18 / 27 | 0.23 | 3.8pp | yes |
| Qwen2.5-Coder-32B dense | 98 | +2.0pp | 4 / 2 | 0.69 | 5.7pp | no |
| agentic BM25 (opus-4-8) | 100 | −2.0pp | 2 / 4 | 0.69 | 5.6pp | no |
| agentic dense (opus-4-8) | 100 | −2.0pp | 2 / 4 | 0.69 | 5.6pp | no |

The two n=499 agentic arms are adequately powered; the n=100 cells and the local Qwen2.5 contrast are underpowered and bound little.

Fix locality. The effect concentrates on localized fixes, which dominate the benchmark: 86% of gold patches touch a single file. Analyzing each of the four single-shot arms separately and combining them with a random-effects meta-analysis (issue as the cluster, arms not pooled as independent), the single-file depth advantage is +6.9pp (95% CI [+5.0, +8.8], significant in each arm, McNemar $p \le 0.008$) and the multi-file stratum is a null +2.2pp (95% CI [−0.7, +5.1], $p=0.13$). The earlier pooled figure (+7.1pp, n=1712) treated the same issue across four arms as independent and is withdrawn. The direction stays positive in every stratum; breadth never significantly outperforms depth. The multi-file case is an underpowered null, not support, and cross-file workloads are required future evidence.

Generality envelope. The core is one benchmark (SWE-bench Verified, popular public Python), one budget (12 slots, about 1.5k tokens under our chunker), one embedder, one chunking scheme. Two of these edges are now mapped rather than open. A second benchmark and three further languages are delivered as a pre-registered boundary (SWE-PolyBench, Java/JavaScript/Python/TypeScript, Section 5.10): the pooled multilingual effect is positive but not significant (+2.6pp, p=0.056), so multilingual generality is a mapped boundary, not a confirmation, and its cross-language heterogeneity tracks the instance pool rather than the surface language. A budget sweep is delivered on a clean model (Qwen3.6-27B at K=4 and K=40, Section 5.9): within the deployable slot range the depth advantage grows with budget (+4.8pp to +9.2pp, within-model difference-in-differences p=0.0003), and any breadth-favoring crossover necessarily lies well beyond K=40, outside the regime we test (Section 3.3). A second embedder, a tools-on seed-context bound, a chunk-size-by-dedup study, and a budget-matched small-to-big baseline remain identified future work, not claims. Every headline is scoped to a 12-slot budget under our chunker.

### 6.3 Practical implications (scoped)

For practitioners serving fixed-slot code context, including pack-style RAG serving, seed contexts for constrained agents, cost-capped batch pipelines, and local models that cannot run agentic loops, the operational recommendation is: do not hard-deduplicate by file at a tight slot budget, and A/B packing policies against the task rather than against the metric the flag was tuned to. The recommendation is operational and needs no oracle; the anchor-dose diagnostic is retrospective and is offered as an explanation, not as an inference-time objective. The evidence supports this for embedding retrievers (fusion and dense) whose depth arm raises anchor dose, on single-file-dominated Python workloads, single-shot. It does not extend to a non-embedding lexical retriever (BM25 reverses), to tool-using search agents (the effect dissolves), or to large-budget regimes far beyond K=40 (where DeepDiscovery reports breadth recovery); across languages it is a mapped boundary rather than a confirmed recommendation (the pooled multilingual test is positive but not significant, Section 5.10); and it is not established for retrievers, embedders, or models not tested at power.

---

## 7. Conclusion

At a tight fixed context budget, the file-deduplication flag adopted for its file-recall@k gain reduces single-shot issue resolution on two confirmatory model arms (gpt-5.6-sol +7.6pp, and a pre-registered open-weights replication Qwen3.6-27B +3.6pp, both surviving a conditional estimand and repository clustering), with a supportive producibility-channel arm (DeepSeek-v4 +6.2pp) and a dense replication that reproduces the direction but whose clustered interval includes zero (DENSE-1 +5.7pp). The effect holds across a roughly 5x resolve-capability span (weak-open A3B +4.2pp, strong-open Qwen3.6-27B +3.6pp, frontier-closed gpt +7.6pp at K=12), grows with the slot budget on a clean model (Qwen3.6-27B +4.8pp at K=4 to +9.2pp at K=40, within-model difference-in-differences p=0.0003), and the open-weights arms can be independently re-run. We frame this as a code-repair instance of the known relevance-diversity, granularity, and objective-mismatch tradeoff rather than a new phenomenon. Its value is threefold: it moves the phenomenon from QA (Levy et al., 2025, itself a controlled manipulation) to execution-graded repository repair driven by a standard recall-tuned packing flag; a random-chunk control refutes the argmax chunk-selection alternative (though it does not by itself separate depth from the file-count change); and the effect comes with a mechanism-linked map of where it appears. The effect holds for the two embedding retrievers tested, does not extend to a lexical BM25 configuration where it reverses (a significant interaction), tracks within-file anchor dose (a supported but partial mediator, carried through a quality channel that is 68 to 94% of every arm's gain, and small under BM25), concentrates on single-file fixes, and is not detected under unrestricted-Read agents. Across four languages on SWE-PolyBench the pre-registered pooled effect is positive but not significant (+2.6pp, p=0.056), a mapped boundary whose cross-language heterogeneity tracks the instance pool rather than the surface language. The BM25 reversal shows the effect is not a universal document-count law, though we do not claim a clean retrieval-paradigm boundary from one configuration. The operational moral is scoped and testable: for fixed-budget packed-context consumers on this kind of workload, do not hard-deduplicate by file, and validate packing policies against the task rather than against the metric the flag was tuned to.

We release the harnesses, predictions, raw responses (the original gpt headline run's responses were not logged; a pre-registered logged replication supplies the gpt parse-apply-test funnel, Section 5.7), grading reports, served packs, pre-registration records, and analysis outputs with a per-instance manifest (Appendix A/B). The artifact is archived at Zenodo, DOI 10.5281/zenodo.21879550 (this version; the concept DOI 10.5281/zenodo.21879549 resolves to the latest version), with the archive `recall-trap-artifact.tar.gz` pinned by SHA-256 `2587f63a98f56738127a84268c134299b539c9c4cd3aeeac3717e037553b0d50`.

---

---

## Appendix A: Reproducibility handles

The released artifact contains the single-shot runner (SEARCH/REPLACE applied to a per-arm isolated checkout, then git diff; raw responses persisted as an audit trail where logged), the agentic harness used only for the boundary condition, the ragd retrieval fleet configuration (fusion backend, `lex_mode=ids`, `w_graph=1.0`, embedder Qwen3-Embedding-8B-Q8_0 via llama.cpp, full un-elided identity string, chunker size/overlap/boundary policy, and the exact `ragd` commit), the served packs for every arm, the predictions, raw responses, grading reports, the per-instance inclusion table, and the analysis scripts. Every table row maps to its inputs through a per-instance manifest. The full artifact is archived at Zenodo (DOI 10.5281/zenodo.21879550, this version; SHA-256 of `recall-trap-artifact.tar.gz` = `2587f63a98f56738127a84268c134299b539c9c4cd3aeeac3717e037553b0d50`); its `MINT-MANIFEST.md` gives the file inventory and the SHA-256 of the verifiable-core files (paper, all pre-registration documents, ledger, harnesses, analysis scripts). Model slugs, providers, and decoding parameters are recorded in the harness configuration and the pre-registration records; per-instance checkouts and Docker images are excluded by design and rebuild from the served packs.

## Appendix B: Pre-registration records

**Provenance classes.** We label every arm by the strongest provenance it actually attains, verifiable from this repository's history. **Class 1, pre-registered:** the design document was committed before any of the arm's data existed. These are `QWEN36-PREREG.md` (commit adc864d, 2026-07-17; the open-weights replication, with Amendment 1 of 2026-07-17 raising the reasoning completion budget 16k→32k, sized blind to the ON/OFF contrast, Section 4.3); `GENERALITY-BENCH-PREREG.md` (commit e3e1ef0; rules G-1–G-5, with the exclusion list and cause classes frozen and committed at b9e36b0 before any A/B resolve); `AGENTIC-BENCH-PREREG-2026-07-15.md` (commit 211261a, 2026-07-16); `CLEAN-CONTROL-DESIGN.md` (commit 470a1c7, 2026-07-15); and the CONTEXT-VAULT §8 registrations (committed 2026-07-14), which fixed, before the powered BM25 run, an exploratory n=100 BM25 pilot showing the same reversal direction and the direction rule for the retriever-agnosticism test. `GPT-LOGGED-REPLICATION-PREREG.md` (commit c35850a, 2026-08-08) is a further Class-1 registration: a fully logged confirmatory replication of the primary gpt-5.6-sol single-shot dedup contrast on the same served packs, with its reporting rule, endpoint hierarchy, and failure branch (clause d) fixed pre-data; its Class-1 label attaches to the replication run and its funnel only, and the original 2026-07-14 headline run remains design-committed and is not upgraded by it (Section 5.7). The wider-program documents are all Class 1 with verified pre-data commits: oracle-ceiling (eae0aeb, 2026-07-14), missing-cell/E0 (dbc1289, 2026-07-14), delivery-mode C1 (8b26d65, 2026-07-14), and g-plane (b3b693d, 2026-07-13); their hashes are in the artifact manifest and the pre-registration ledger. **Class 2, design notes without a pre-data commit:** `G2-RANDOM-CHUNK-PREREG.md` (the random-chunk decision rule) and `BM25-AGNOSTIC-PREREG.md` (the TOST margin and bound-reporting commitment) were authored as dated working notes but not committed before their runs; they are committed at release with provenance headers, a file date is not a cryptographic anchor, and no arm governed only by a Class-2 note is called pre-registered in this paper.

**Timing disclosures.** The E1–E2 allocation design (commit 1eaab36) was committed on 2026-07-14, the same calendar day the primary gpt-5.6-sol result was first unblinded; the within-day design-before-data ordering rests on released run logs, not on git alone, so gpt and DeepSeek are described as design-committed confirmatory arms rather than pre-registered ones; gpt additionally carries a Class-1 pre-registered logged confirmatory replication of its direction and funnel (GPT-LOGGED-REPLICATION-PREREG, Section 5.7), which upgrades the replication and the funnel, not the original headline estimate. DENSE-1 (labeled "CC-1", clean-control-1, in the committed design docs, harness, and released artifact, renamed here only to avoid collision with the Chroma Context-1 system) is deliberately not listed as a pre-registered arm: its dense-control harness (commit c7face1) and its analysis-design note were committed on 2026-07-15, about a day after the primary gpt-5.6-sol result was first unblinded (2026-07-14), though before DENSE-1's own run later that evening. DENSE-1 is therefore reported as a replication and generality control specified before its own data but after the primary effect, not as a control registered ahead of the primary result (Section 5.3). The budget×dedup dose-response (Section 5.9) has no pre-registration document; an earlier draft referred to a "budget-sweep pre-registration", no such committed document exists, and the reference is withdrawn. The word "pre-registered" is used only for Class-1 arms; where an ordering is same-day and rests on logs, we say "design-committed" instead. Every Class-1 commit hash above is verified and mirrored in `PREREG-LEDGER.md`; the artifact is archived at Zenodo (DOI 10.5281/zenodo.21879550, this version; SHA-256 `2587f63a98f56738127a84268c134299b539c9c4cd3aeeac3717e037553b0d50`).

## Appendix C: Reasoning-model incompatibility log

GLM-5.2 and Kimi-k2.7-code are documented as incompatible with the single-shot no-tools regime (provider-side unbounded or mandatory reasoning under a bounded completion budget), with raw-response evidence, and counted as neither replications nor non-replications.

## Appendix D: Hypothesis registry

| # | hypothesis | pre-reg ref | data used | disposition | type |
|---|---|---|---|---|---|
| 1 | oracle ceiling (perfect pack beats explorer) | oracle-ceiling | n=92 | +10.9pp, real but anticipated by SWE-Explore (2606.07297) | not-novel |
| 2 | tools actively harm a well-contextualized agent | e0-missing-cell | E0 (oracle+tools) | +2.2pp, p=0.63; inconclusive (underpowered), read as no support | inconclusive |
| 3 | delivery mode: pre-inject beats tool-call | c1-delivery-mode | n=92 then n=499 | +8.7pp@92 collapses to +1.6pp p=0.45@499 | inconclusive-at-scale / not-novel |
| 4 | right-file-wrong-lines as a novel failure mode | — | §5.7 | anticipated by RGFL (Sepidband et al., 2026) | not-novel |
| 5 | smaller pack beats bigger as novel | — | — | anticipated by the SWE-bench retrieval-context ablation (Jimenez et al., 2024, 2310.06770, App. A) | not-novel |
| 6 | planner harness / small-model retrieval driver | — | — | anticipated (Chroma Context-1, FastContext, CoSIL) | not-novel |
| 7 | file-dedup costs resolution at fixed budget | e1-e2-allocation (committed 2026-07-14; same-day timing, App B) + QWEN36-prereg; DENSE-1 post-primary (App B) | §5 | survivor | supported |
| 8 | agentic +18.2pp as a robustness result | — | n=499 | withdrawn: model confound (sonnet vs opus), Section 5.5 | confounded |
| 9 | the effect is fusion-specific | — | §5.3 | withdrawn in direction: DENSE-1 dense reproduces it (instance-level p=0.0097; clustered CI includes zero) | superseded |
| 10 | argmax-per-file selection is the harm (keeps the wrong chunk) | — (Class-2 design note, no pre-data commit; App B) | §5.4 | refuted: random reselection is worse than argmax (Qwen B 3.0% vs A 6.6%) | refuted |
| 11 | the effect is retriever-agnostic (extends to lexical BM25) | CONTEXT-VAULT §8 (n=100 pilot + direction rule, committed 2026-07-14); TOST design = Class-2 note (App B) | §5.3 | refuted: BM25 reverses (−3.2pp), significant interaction; scope limited to the two embedding retrievers tested | refuted |
| 12 | the effect reproduces on | QWEN36-prereg | §5.1 | Qwen3.6-27B +3.6pp, p=0.0133, survives clustering ([+0.9,+4.9]pp, all LORO folds positive) | supported (confirmatory) |

| # | hypothesis | pre-reg ref | data used | disposition | type |
|---|---|---|---|---|---|
| | an open-weights model | (+Amendment 1) | | | |
| 13 | budget×dedup dose-response (the trap grows with slot budget K) | — (no prereg; exploratory sweep identified the clean confirming runs, App B) | §5.9 | clean Qwen3.6-27B +4.8pp@K=4 → +9.2pp@K=40, within-model difference-in-differences p=0.0003; weak-open A3B +2.0/+4.2/+6.2pp independently confirms the direction (exploratory) | supported (exploratory→replicat |
| 14 | multilingual generality (the trap holds across languages) | GENERALITY-BENCH-prereg (G-1–G-5) | §5.10 | pooled two-model +2.6pp, McNemar p=0.056 (G-4 branch); positive but not significant, no cell survives Holm, heterogeneity tracks instance pool not language | mapped-boundary (nc supported) |
| 15 | the primary gpt effect replicates under full response logging, and test-pass-conditional-on-applied favors depth (funnel primary) | GPT-LOGGED-REPLICATION-prereg (clause d) | §5.7 | replication Δ=+6.4pp (McNemar p=0.002, clustered CI [+4.9,+11.3]); funnel dedup-OFF>ON on test-pass\|applied (+5.2pp); direction confirmed, downgrade branch did not fire | supported (confirmatory) |

The registry has fifteen entries. "The four single-shot runs" in entry 7 are the gpt, Qwen3.6-27B, DeepSeek, and DENSE-1 A/Bs of Section 5.1; their design provenance is heterogeneous and stated arm-by-arm in Appendix B. Qwen3.6-27B is pre-registered (design committed before any of its data). gpt and DeepSeek are design-committed confirmatory arms: their governing E1–E2 allocation design was committed on 2026-07-14, the same day the primary result was first unblinded, so the within-day design-before-data ordering rests on released run logs rather than on git alone. DENSE-1 was specified before its own data but after the primary result. Entries 10 and 11 are the two mechanism/scope controls added this version; entry 11's pilot and direction rule carry a 2026-07-14 commit while its TOST design is a Class-2 note, and entry 10 has no pre-data commit (Appendix B). Dispositions are of distinct types (not-novel, refuted, inconclusive-underpowered, confounded, superseded, supported, and mapped-boundary); "died" is not a single kind of event, and we relabel the underpowered nulls (entries 2, 3) as inconclusive rather than refuted. Entry 14 (multilingual generality) is a pre-registered result that landed on the G-4 honest-boundary branch: directionally positive but below the pre-registered significance threshold, reported as a mapped edge rather than a confirmation. Entry 15 (gpt logged replication) is a pre-registered confirmatory replication whose reporting rule and downgrade branch were fixed pre-data (Appendix B); it reproduced the direction (+6.4pp, inside the original clustered CI) and supplied the gpt parse-apply-test funnel, and its downgrade branch did not fire.